\documentclass{acm_proc_article-sp}
\usepackage{url}
\usepackage{multirow}
\usepackage{balance}
\usepackage{rotating}
\usepackage{booktabs}
\usepackage{hyphsubst}
\usepackage{graphicx}
\usepackage{amsmath}
\graphicspath{{images/}}
\usepackage{hhline}
\usepackage[caption = false]{subfig}

\usepackage{url}
\usepackage{rotating}
\usepackage{amsmath}
\usepackage{cancel}
\usepackage{amssymb}
\usepackage{enumitem}
\usepackage{pifont}

\usepackage{multirow}
\usepackage{array}
\usepackage{placeins}

\usepackage{rotating}
\usepackage{booktabs}
\usepackage{rotating}
\usepackage{hhline}
\usepackage{hyphsubst}
\usepackage{soul}

\permission{Permission to make digital or hard copies of all or part of this work for personal or classroom use is granted without fee provided that copies are not made or distributed for profit or commercial advantage and that copies bear this notice and the full citation on the first page. Copyrights for components of this work owned by others than ACM must be honored. Abstracting with credit is permitted. To copy otherwise, or republish, to post on servers or to redistribute to lists, requires prior specific permission and/or a fee. Request permissions from permissions@acm.org.}
\conferenceinfo{HT'14,}{September 1--4, 2014, Santiago, Chile.}
\copyrightetc{Copyright 2014 ACM \the\acmcopyr}
\crdata{xxx-x-xxxx-xxxx-x/xx/xx\ ...\$xx.xx}

\newlength{\Oldarrayrulewidth}

\begin{document}

\title{Recommending Items in Social Tagging Systems Using Tag and Time Information}


\numberofauthors{6} 
\author{
\alignauthor
Emanuel Lacic\\
\affaddr{Knowledge Technology Institute}\\
       \affaddr{Graz University of Technology}\\
       \affaddr{Graz, Austria}\\
       \email{elacic@know-center.at}
\alignauthor
Dominik Kowald\\
\affaddr{Know-Center}\\
       \affaddr{Graz University of Technology}\\
       \affaddr{Graz, Austria}\\
       \email{dkowald@know-center.at}
       \and
\alignauthor
Paul Seitlinger\\
       \affaddr{Knowledge Technology Institute}\\
       \affaddr{Graz University of Technology}\\
       \affaddr{Graz, Austria}\\
       \email{paul.seitlinger@tugraz.at}
\alignauthor 
Christoph Trattner\\
\affaddr{Know-Center}\\
       \affaddr{Graz University of Technology}\\
       \affaddr{Graz, Austria}\\
       \email{ctrattner@know-center.at}
\alignauthor 
Denis Parra\\
\affaddr{CS Department}\\
\affaddr{Pontificia Universidad Cat\'olica de Chile}\\
       \affaddr{Santiago, Chile}\\
       \email{dparra@ing.puc.cl}
}

\maketitle

\begin{abstract}
In this work we present a novel item recommendation approach that aims at improving Collaborative Filtering (CF) in social tagging systems using the information about tags and time. 
Our algorithm follows a two-step approach, where in the first step a potentially interesting candidate item-set is found using user-based CF and in the second step this candidate item-set is ranked using item-based CF. Within this ranking step we integrate the information of tag usage and time using the Base-Level Learning (BLL) equation coming from human memory theory that is used to determine the reuse-probability of words and tags using a power-law forgetting function.

As the results of our extensive evaluation conducted on data-sets gathered from three social tagging systems (BibSonomy, CiteULike and MovieLens) show, the usage of tag-based and time 
information via the BLL equation also helps to improve the ranking and recommendation process of items and thus, can be used to realize an effective item recommender that outperforms two 
alternative algorithms which also exploit time and tag-based information.

\end{abstract}

\section{Introduction} \label{sec:introduction}
Over the past few years social tagging gained tremendously in popularity, helping people for instance to categorize or describe resources on the Web for better information retrieval (e.g., BibSonomy or CiteULike) \cite{korner2010stop, Trattner2012}. Although the process of tagging has been well explored in the past and in particular the task of predicting the right tags to the user in a personalized manner \cite{jaschke2008tag, Seitlinger2013}, studies on predictive models to recommend items to users based on social tags are still rare. To contribute to this sparse field of research, in this paper we present preliminary results of a study that aims at addressing this issue. In particular, we provide first results of a novel attempt to improve item recommendations by taking into account peoples' social tags and the information of the time the tags have been applied by the users. As shown in related work, recommending items to users in a collaborative manner relying on social tagging information is not an easy task in general (e.g., \cite{tso2008tag} or \cite{parra2010improving}). However, other related work has also proofed that the information of time is an important factor to make the models more accurate in the end (e.g., \cite{Zheng2011} or \cite{Huang2014}). 

Contrary to the previous work mentioned above, we suggest a less data-driven approach that is inspired by principles of human memory theory about remembering things over time. As shown in our previous work on tag recommender systems \cite{domi2014}, the base-level learning (BLL) equation introduced by Anderson and Schooler \cite{Schooler1991} (see also Anderson et al. \cite{Anderson2004}), which integrates tag frequency and recency (i.e., the time since the last tag usage), can be used to implement an effective tag recommendation and ranking algorithm. In particular, the BLL equation models the time-depended drift of forgetting of words and tags using a power-law distribution in order to determine a probability value that a specific tag will be reused by a target user.

In this work, we apply this equation for ranking and recommending items to users. To this end, we present a novel recommender approach called \textit{Collaborative Item Ranking Using Tag and Time Information (CIRTT)} that firstly identifies a potentially interesting candidate item set and secondly, ranks this candidate set in a personalized manner (similar to \cite{Huang2014}). In this second step of personalization, we integrate the BLL equation to include this information about tags and time. To investigate the question as to whether tag and time information can improve the ranking and recommendation process, we conducted an extensive evaluation using folksonomy datasets gathered from three social tagging systems (BibSonomy, CiteULike and MovieLens). Within this study we compared our approach to two alternative tag and time based recommender algorithms \cite{Zheng2011,Huang2014} amongst others. The results show that integrating tag and time information using the BLL equation helps to improve item recommendations and to outperform state-of-the-art baselines in terms of recommender accuracy.

The remainder of this paper is organized as follows. We begin with explaining our tag and time based approach CIRTT in Section \ref{sec:approach}. Then we describe the experimental setup of our evaluation in Section \ref{sec:expset} and summarize the results of this study in Section \ref{sec:results}. Finally, in Section \ref{sec:con}, we close the paper with a short conclusion and an outlook into the future.


\section{Approach} \label{sec:approach}

In this section we provide a detailed description of our item recommendation approach called \textit{Collaborative Item Ranking Using Tag and Time Information (CIRTT)}. 
In general, our CIRTT algorithm uses a similar strategy as the approach proposed by Huang et al. \cite{Huang2014} and thus, consists of two steps relying on a combination of user- and item-based CF: in the first step, a potentially interesting candidate item set for the target user \textit{u} is determined and in the second step, this candidate item set gets ranked using item similarities and tag and time information.

Step one (i.e., determining candidate items) is conducted using a simple user-based CF approach. Hence, we first find the most similar users for the target user \textit{u} (i.e., the neighborhood) based on the binary user-item matrix $B_{u,i}$ (see also \cite{Zheng2011}) and then, use the bookmarked items of these neighbours as our candidate item set. We use a neighbourhood of \textit{k} = 20 users and the Cosine similarity measure \cite{gemmell2009improving} (see also Section \ref{sec:baselines}).

In the second step (i.e., ranking candidate items) we use an item-based CF approach in order to determine the relevance of each candidate item for the target user based on the items she has bookmarked in the past. Hence, for each candidate item \textit{i} in the candidate item set we calculate this combined similarity value $sim(u, i)$ by the item-based CF formula: 
\begin{align}
		sim(u, i) = \sum\limits_{j \in items(u)}{sim(i, j)}
\end{align}
, where $items(u)$ is the set of items the target user $u$ has bookmarked in the past. This item-based CF step helps us to give a higher ranking to candidate items that are more similar to the items the target user has bookmarked in the past (see also \cite{Huang2014}).


To finally realize CIRTT in order to integrate tag and time information we make use of the base-level learning (BLL) equation proposed by Anderson et al. \cite{Anderson2004}. As described in our previous work \cite{domi2014}, the BLL equation can be used to determine a relevance value for a tag $t$ in the tag assignments of a target user $u$ based on tag frequency and recency:
	\begin{align}
		BLL(u, t) = ln(\sum\limits_{i = 1}\limits^n{t_{i}^{-d})}
  \end{align}
, where $n$ is the number of times $t$ has been used by $u$ and $t_{i}$ is the recency, i.e., the time since the $i^{th}$ occurrence of $t$ in the tag assignments of $u$. The exponent $d$ is used to model the power law of forgetting memory items and is usually set to $.5$ (see \cite{Anderson2004}). In order to map these BLL values on a range of 0 - 1, we used the same normalization method as used in our previous work \cite{domi2014}.

\begin{table}[t!]
  \setlength{\tabcolsep}{2.6pt}
  \centering
    \begin{tabular}{l|lllll}
    \specialrule{.2em}{.1em}{.1em}
      Dataset			& $|B|$			& $|U|$	& $|R|$	& $|T|$	& $|TAS|$	 \\ \hline 
      BibSonomy	 	& 82,539 	& 2,437  	& 28,000		& 30,919		& 339,337						\\\hline

			CiteULike		& 36,471  & 3,202  	& 15,400		& 20,937		& 99,635							\\\hline

			MovieLens		& 53,607 	& 3,983 	& 5,724			& 14,883		& 92,387							\\
																		
		\specialrule{.2em}{.1em}{.1em}								
    \end{tabular}
    \caption{Properties of the datasets, where $|B|$ is the number of bookmarks, $|U|$ the number of users, $|R|$ the number of resources, $|T|$ the number of tags and $|TAS|$ the number of tag assignments.}
  \label{tab:dataset_stats}
\end{table}

We adopt this equation for the ranking of items in social tagging systems using a similar method as proposed in \cite{Zheng2011} and \cite{Huang2014}. Thus, a user is assumed to prefer an item if it has been tagged with tags of high relevance for the user, that is, with tags exhibiting a high BLL value. Given this assumption, the BLL value of a given item $i$ for the target user $u$ is determined using the following formula:
	\begin{align}
		BLL(u, i) = \sum\limits_{t \cup tags(u, i)}{BLL(u, t)}
  \end{align}
, where $tags(u, i)$ is the set of tags $u$ has used to tag $i$.

Taken together, the prediction value $pred(u, i)$ of a candidate item $i$ using our CIRTT approach is given by:
\begin{align}
		pred(u, i) = \underbrace{\sum\limits_{j \in items(u)}{sim(i, j)}}_{sim(u, i)} \times BLL(u, i)
\end{align}
This approach enables us to weight higher the items within the candidate set that are more important to the target user (i.e., items associated with tags exhibiting a high BLL value that integrates tag frequency and recency). CIRTT and the baseline algorithms presented in this work are implemented in the Java programming language, are open-source software and can be downloaded online from our Github Repository\footnote{https://github.com/learning-layers/TagRec/} \cite{Kowald2014a}.

\section{Experimental Setup}
\label{sec:expset}

In this section we describe in detail the datasets, the evaluation methodology and metrics as well as the baseline algorithms used for our experiments.

\subsection{Datasets} \label{sec:datasets}
In order to evaluate our approach and for reasons of reproducibility we used freely-available folksonomies gathered from three well-known social-tagging systems. We used data-sets of the social bookmark and publication sharing system BibSonomy\footnote{\url{http://www.kde.cs.uni-kassel.de/bibsonomy/dumps}}, the reference management system CiteULike\footnote{\url{http://www.citeulike.org/faq/data.adp}} and the movie recommendation site MovieLens\footnote{\url{http://grouplens.org/datasets/movielens/}}. As suggested by related work in the field (e.g. \cite{jaschke2007tag, hotho2006information}), we excluded all automatically imported and generated tags (e.g., bibtex-import). 
In the case of CiteULike we randomly selected 10\% of the user profiles for reasons of computational effort (see also \cite{gemmell2009improving}).

We did not use a full \textit{p}-core pruning technique, since this would negatively influence the recommender evaluation results in social tagging system as shown by Doerfel and J{\"a}schke \cite{doerfel2013analysis}, but excluded all unique resources (i.e., resources that have been bookmarked only by a single user). The final dataset statistics can be found in Table \ref{tab:dataset_stats}.

\subsection{Evaluation Methodology} \label{sec:evalmethod}
To evaluate our item recommender approach we used a training and test-set split method as proposed by popular and related work in this area \cite{Huang2014,Zheng2011}. Hence, for each user we sorted her bookmarks in chronological order and used the 20\% most recent bookmarks for testing and the rest for training. With the training set we examined then whether a recommender approach could predict the bookmarked resources of a target user in the test set. This procedure also simulates well a real environment where the bookmarking behavior of a user in the future is tried to be predicted based on the bookmarking behavior in the past \cite{campos2013time}.

To finally quantify the recommendation accuracy of our approaches, we used a set of well-known information retrieval metrics. In particular, we report Normalized Discounted Cumulative Gain (nDCG@20),  Mean Average Precision (MAP @20), Recall (R@20), Diversity (D) and User Coverage (UC) \cite{SmythMcClave01, herlocker2004evaluating}. All performance metrics are calculated and reported based on the top-20 recommended items. Moreover we also show the performance of the algorithms in the plots of all three accuracy metrics (nDCG, MAP and Recall) for 1 - 20 recommended items (see also \cite{cremonesi2011top}).

\subsection{Baseline Algorithms}  \label{sec:baselines}
In order to evaluate our tag and time based CIRTT approach, we compared it to several baseline algorithms in terms of recommender accuracy. The algorithms have been selected with respect to their popularity, performance and novelty.

\textbf{MostPopular (MP):} The most basic approach we utilized is the simple \textit{Most Popular (MP)} approach that recommends for any user the same set of items. These items are weighted by their frequency in all bookmarks, meaning that the most frequently bookmarked items are recommended.

\textbf{User-based Collaborative Filtering (CF):} Another approach we benchmarked against is the well-known \textit{User-based Collaborative Filtering (CF)} recommendation algorithm \cite{schafer2007collaborative}. The main idea of CF is that users that are more similar to each other (i.e., have similar taste), will probably also like the same items. Thus, the CF approach first finds the $k$ most similar users for the target user and afterwards recommends their items that are new to her (i.e., have not been bookmarked before). We calculated the user-similarities based on both, the binary user-item matrix as proposed in \cite{Zheng2011} (hereinafter referred to as \textit{CF$_B$}) and the tag-based user profiles as proposed in \cite{Huang2014} (hereinafter referred to as \textit{CF$_T$}). Although we also considered using \emph{Item-based CF} \cite{Sarwar2001}, we dismissed it based on the tag-based recommender experiments of Bogers et al. \cite{Bogers2008} showing that user-based CF always beat item-based CF. They explain the result given that the number of items in the dataset is larger than the number of users, and this is also the case in our three datasets (Table \ref{tab:dataset_stats}). 


\begin{table*}[t!]
\begin{center}
\begin{tabular}{l|llllll|l}
\specialrule{.2em}{.1em}{.1em}
  Dataset & Metric
  					&  $MP$ 	& $CF_{T}$ 	& $CF_{B}$ 	& $Z$ & $H$ & 
  					$CIRTT$  \\\hline
  \multirow{5}{*}{\centering{\centering{BibSonomy}}}  
  &  nDCG@20 		&	$.0143$ 	&	$.0448$ 	& $.0610$ & $.0621$ & $.0564$ & 
  $.\textbf{0638}$ \\ 
  &  MAP@20 		&	$.0057$ 	&	$.0319$ 	& $.0440$ & $.0447$ & $.0394$ & 
  $.\textbf{0464}$ \\  
  &  R@20	 		&	$.0204$ 	&	$.0618$ 	& $.0820$ & $.0834$ & $.0816$ & 
  $.\textbf{0907}$ \\  
  &  D 			&	$.8307$ 	&	$.8275$ 	& $.8852$ & $.8528$ & $.6209$ & 
  $.8811$ \\ 
  &  UC 			&	$100\%$ 	&	$99.76\%$ 	& $99.52\%$ 	& $99.52\%$ & $99.76\%$ & 
  $99.76\%$   \\    
    \hline 
  \multirow{5}{*}{\centering{\centering{CiteULike}}}   
  &  nDCG@20 		&	$.0062$ 	&	$.0407$ 	& $.0717$ & $.0762$ & $.0706$ & 
  $.\textbf{0912}$ \\ 
  &  MAP@20 		&	$.0036$ 	&	$.0241$ 	& $.0453$ & $.0484$ & $.0459$ & 
  $.\textbf{0629}$ \\   
  &  R@20 			&	$.0077$ 	&	$.0630$ 	& $.1033$ & $.1077$ & $.0928$ & 
  $.\textbf{1225}$ \\    
  &  D 			&	$.8936$ 	&	$.7969$ 	& $.8642$ & $.8145$ & $.6318$ & 
  $.8640$ \\ 
  &  UC 			&	$100\%$ 	&	$98.38\%$ 	& $96.44\%$ 	& $97.32\%$ & $98.38\%$ & 
  $97.61\%$   \\   
  \hline 
  \multirow{5}{*}{\centering{\centering{MovieLens}}} 
  &  nDCG@20 		&	$.0198$ 	&	$.0361$ 	& $.0602$ & $.0614$ & $.0484$ & 
  $.\textbf{0650}$ \\
  &  MAP@20 		&	$.0075$ 	&	$.0201$ 	& $.0347$ & $.0367$ & $.0263$ & 
  $.\textbf{0413}$ \\    
  &  R@20 			&	$.0366$ 	&	$.0561$ 	& $.1031$ & $.1013$ & $.0763$ & 
  $.\textbf{1058}$ \\  
  &  D 			&	$.9326$ 	&	$.8861$ 	& $.9267$ & $.9119$ & $.7789$ & 
  $.9176$ \\  
  &  UC 			&	$100\%$ 	&	$97.82\%$ 	& $95.90\%$ 	& $98.43\%$ & $97.82\%$ & 
  $95.90\%$   \\
    \specialrule{.2em}{.1em}{.1em}
\end{tabular}
\vspace{-2mm}
\caption{nDCG@20, MAP@20, R@20, D and UC values for BibSonomy, CiteULike and MovieLens showing that CIRTT, that integrates tag and time information using the BLL-equation, outperforms state-of-the-art baseline algorithms.}
 \label{tab:full_norm}
\end{center}
\vspace{-6mm}
\end{table*}

\textbf{Collaborative Filtering Using Tag and Time Information (Z / H):} We also compared our approach to two alternative algorithms that focus on improving Collaborative Filtering for social tagging systems using tag and time information. The first one has been proposed by Zheng et al. \cite{Zheng2011} (hereinafter referred to as \textit{Z}) and improves the traditional CF approach based on the binary user-resource matrix using tag and time information. As in our CIRTT approach this is done using information about tag frequency and recency but in contrast to our solution the authors model the forgetting process using an exponential distribution rather than a power-law distribution. Moreover, this information is already used in the user similarity calculation step and not in the item ranking step as it is done in our approach.


The second tag and time-based approach we tried to benchmark against was proposed by Huang et al. \cite{Huang2014} (hereinafter referred to as \textit{H}). As in our approach, this algorithm uses a 2-step recommendation process, where in the first step a potentially interesting candidate item-set for the target user is determined using user-based CF and in the second step this candidate item-set is ranked using item-based CF. In contrast to our approach, the authors calculate the user and item similarities based on user tag-profiles rather than based on the binary user-item matrix. Furthermore, in this approach the forgetting process is modeled using a simple linear function rather than a power-law distribution.


All CF-based approaches mentioned in this section use a neighborhood of 20 users and make use of the Cosine similarity measure as it is also done in CIRTT (see also \cite{gemmell2009improving}).
\begin{figure*}[ht!]
  \centering
  		 \subfloat[\hspace*{1.8em} nDCG \newline BibSonomy]{ 
				\includegraphics[width=0.33\textwidth]{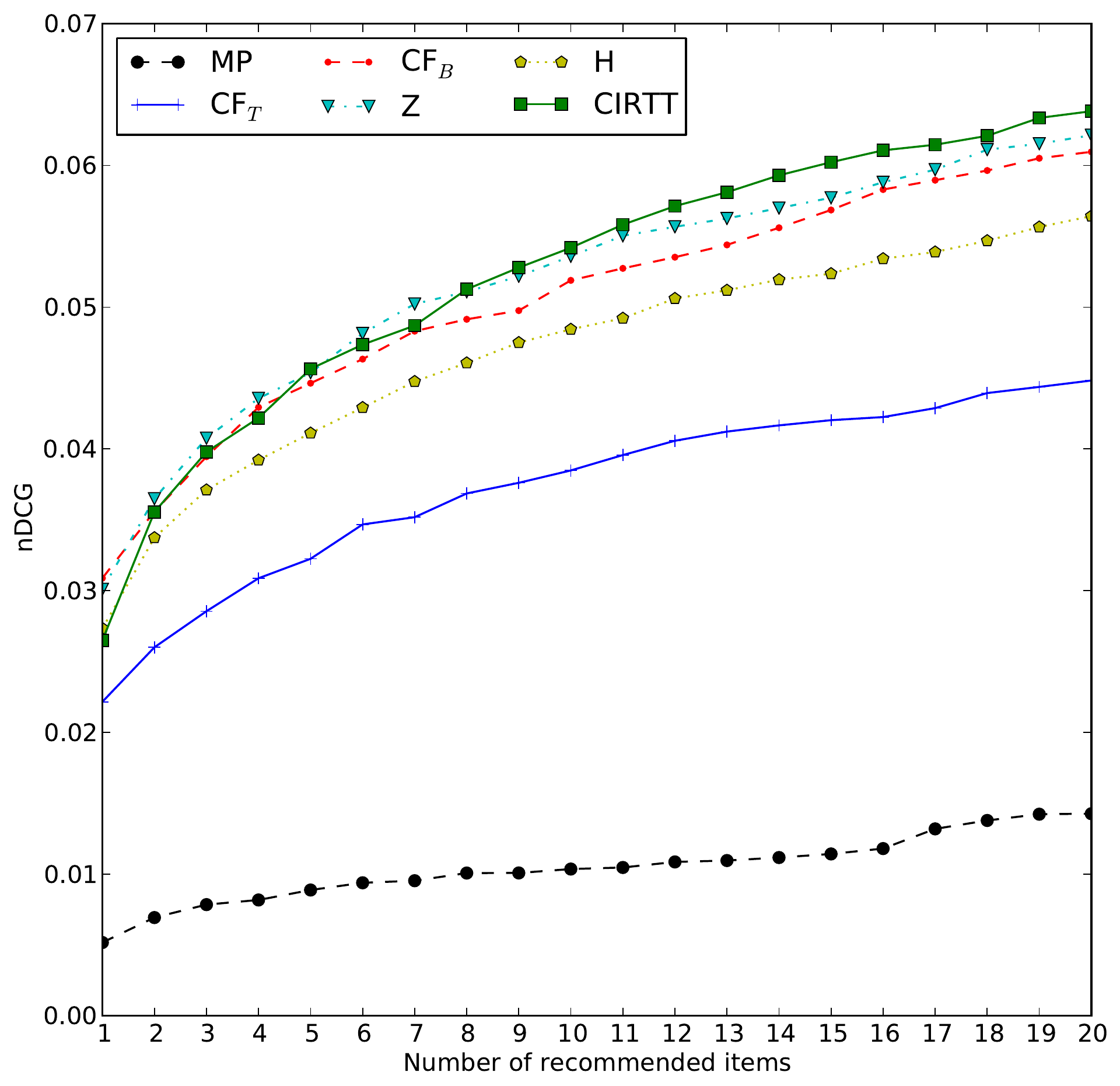}%
		 } 
		 \subfloat[\hspace*{1.8em} nDCG \newline CiteULike]{ 
				\includegraphics[width=0.33\textwidth]{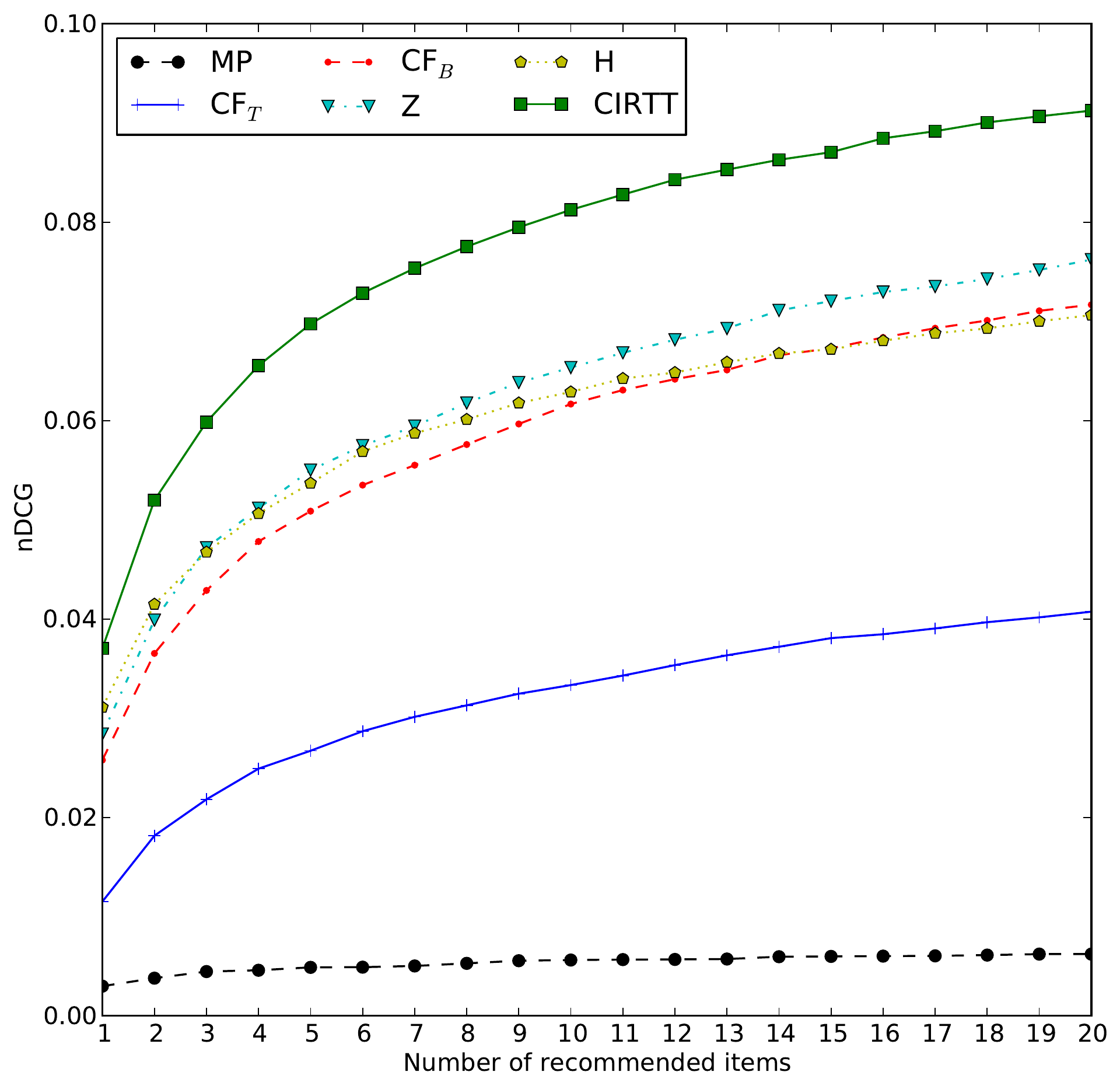}%
		 }
		 \subfloat[\hspace*{1.8em} nDCG \newline MovieLens]{ 
				\includegraphics[width=0.33\textwidth]{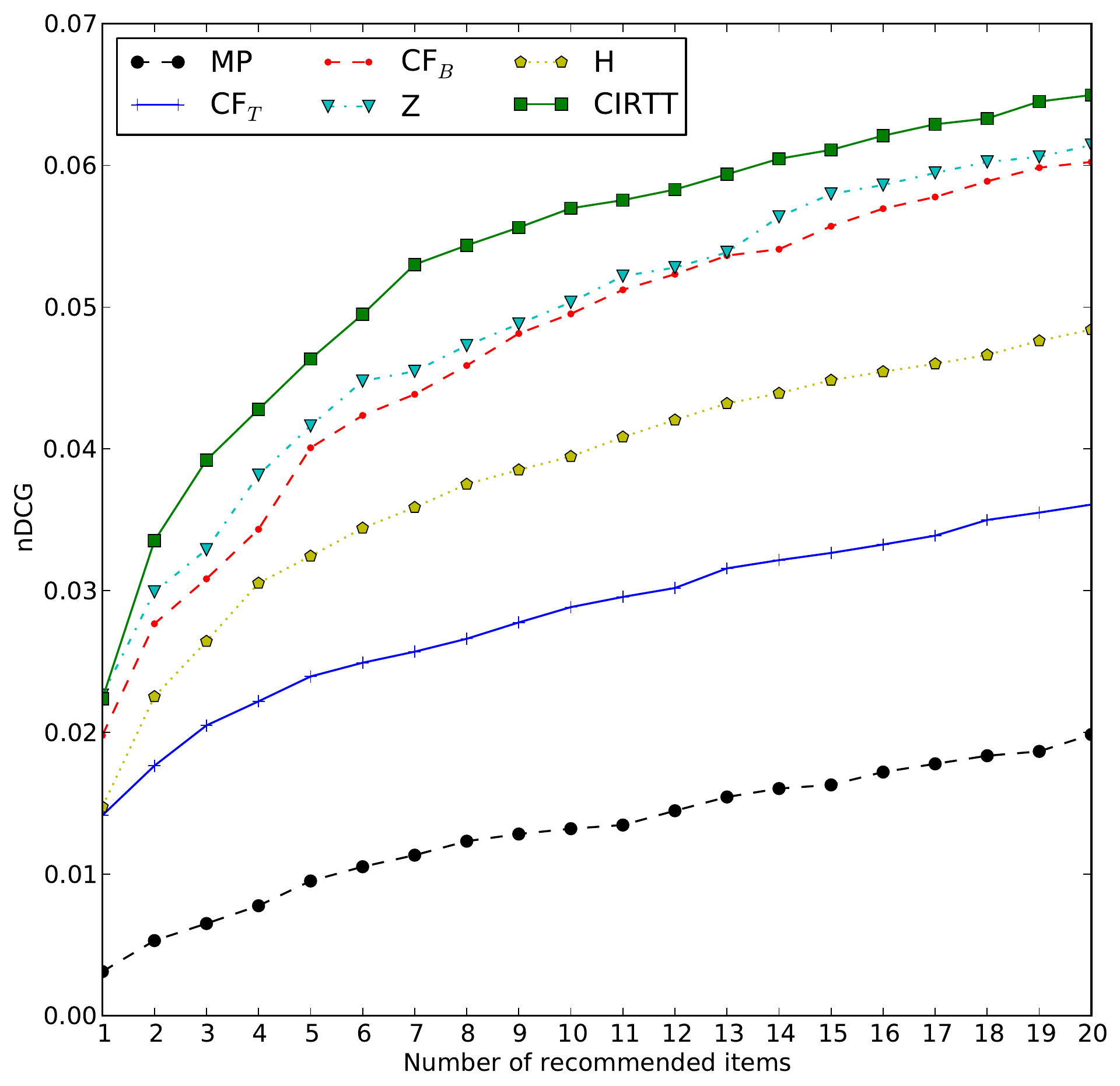}%
		 } \\
		 \subfloat[\hspace*{1.8em} MAP \newline BibSonomy]{ 
				\includegraphics[width=0.33\textwidth]{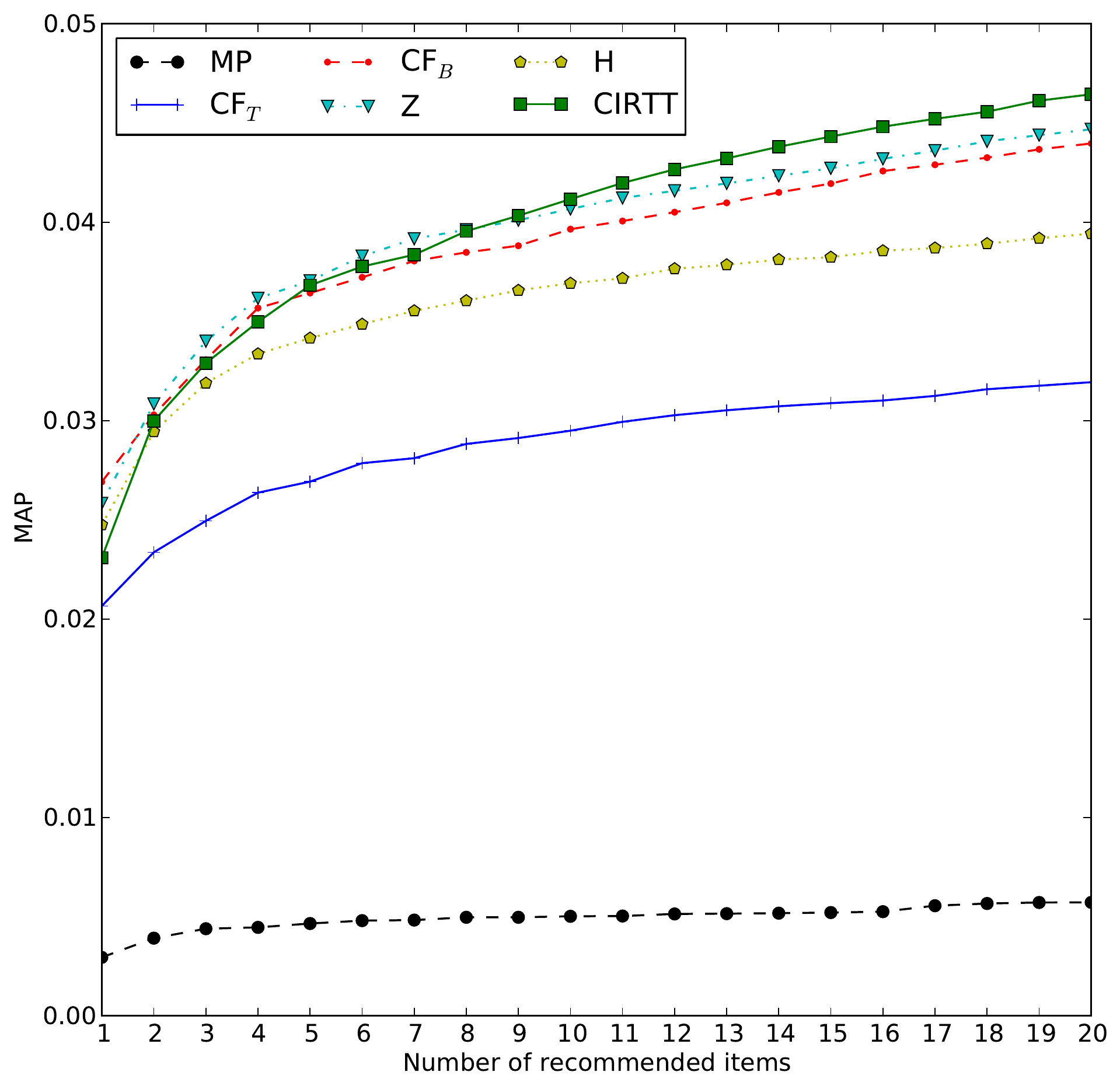}%
		 } 
		 \subfloat[\hspace*{1.8em} MAP \newline CiteULike]{ 
				\includegraphics[width=0.33\textwidth]{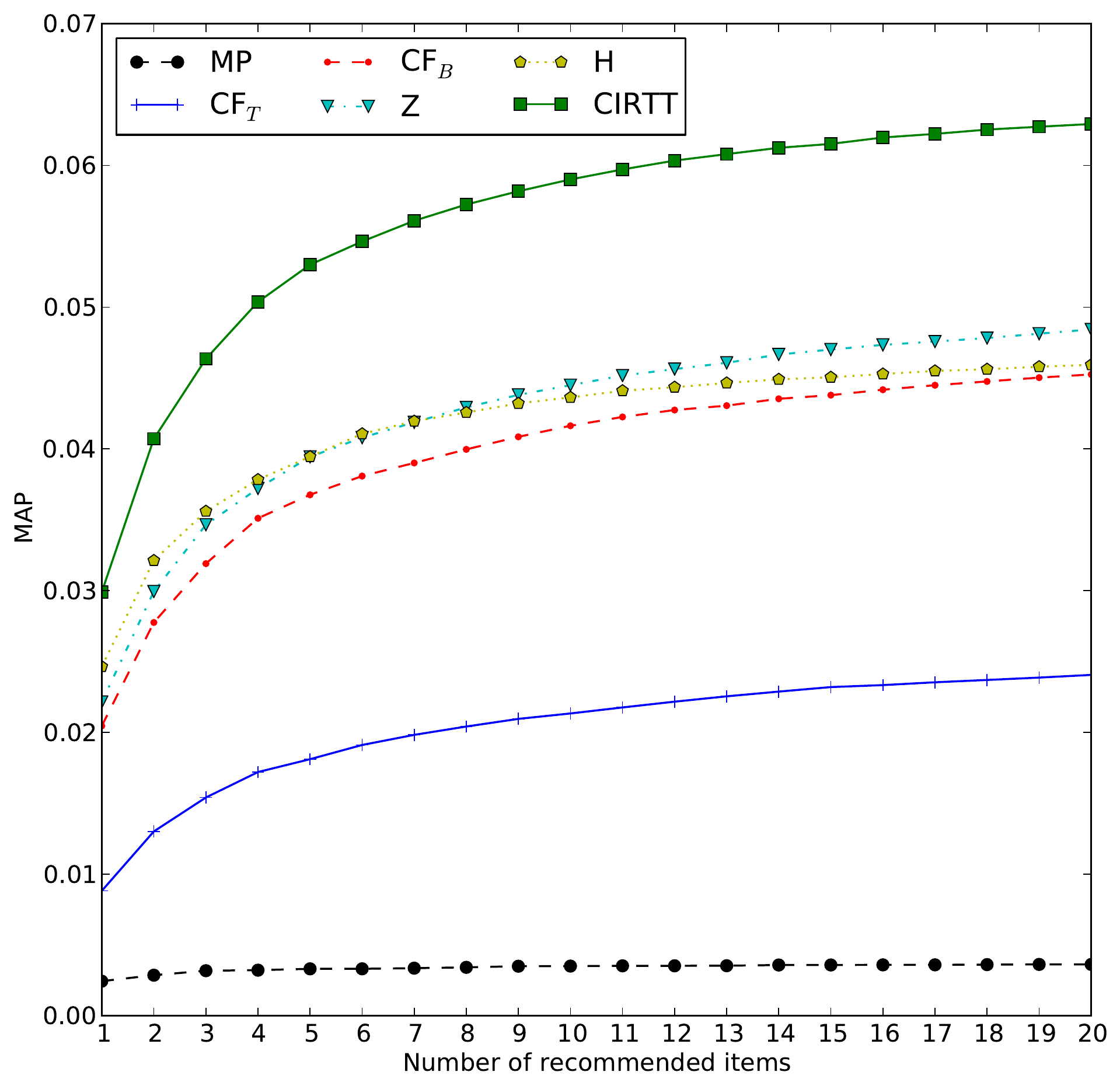}%
		 }
		 \subfloat[\hspace*{1.8em} MAP \newline MovieLens]{ 
				\includegraphics[width=0.33\textwidth]{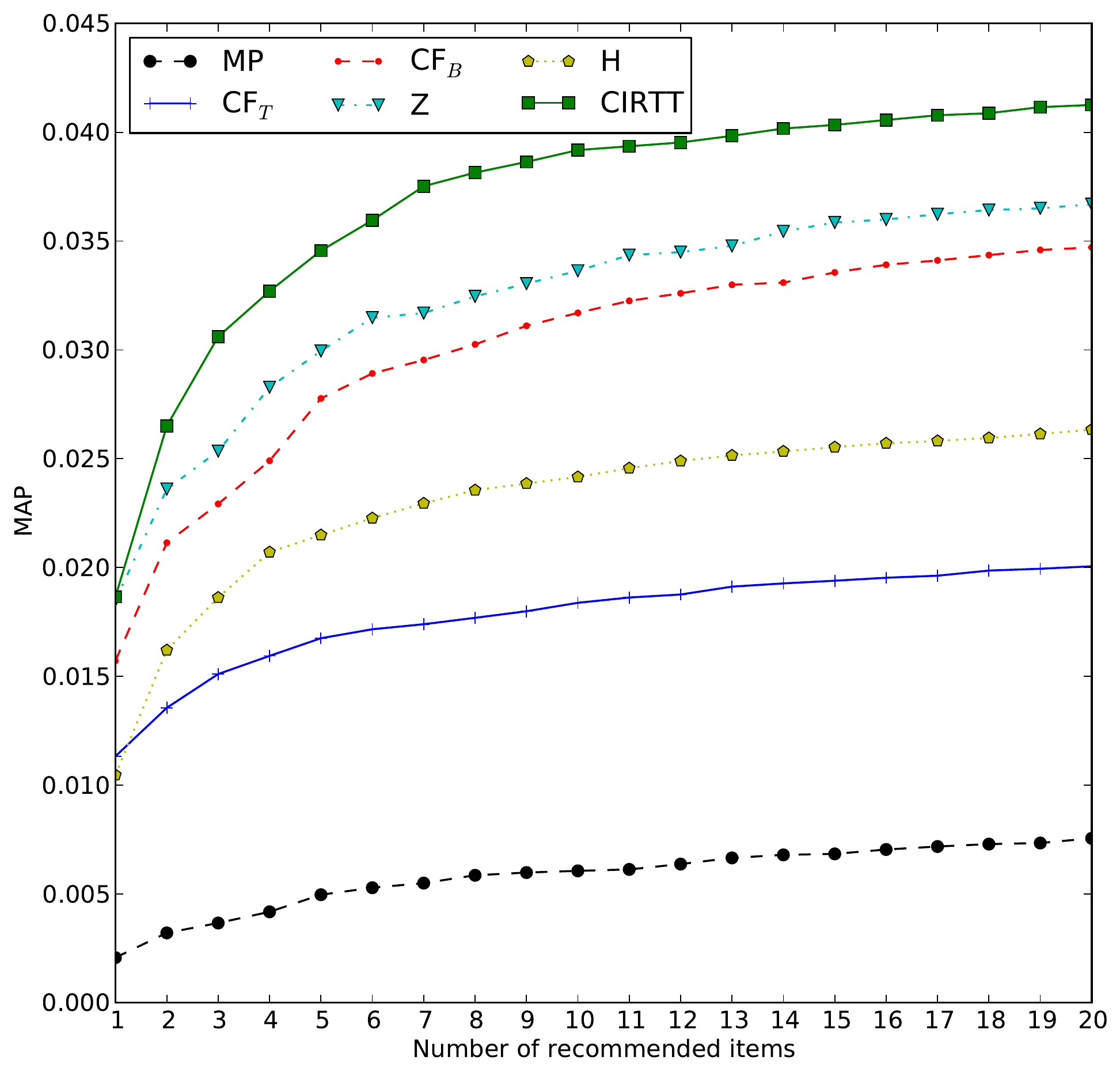}%
		 } \\
		 \subfloat[\hspace*{1.8em} Recall \newline BibSonomy]{ 
				\includegraphics[width=0.33\textwidth]{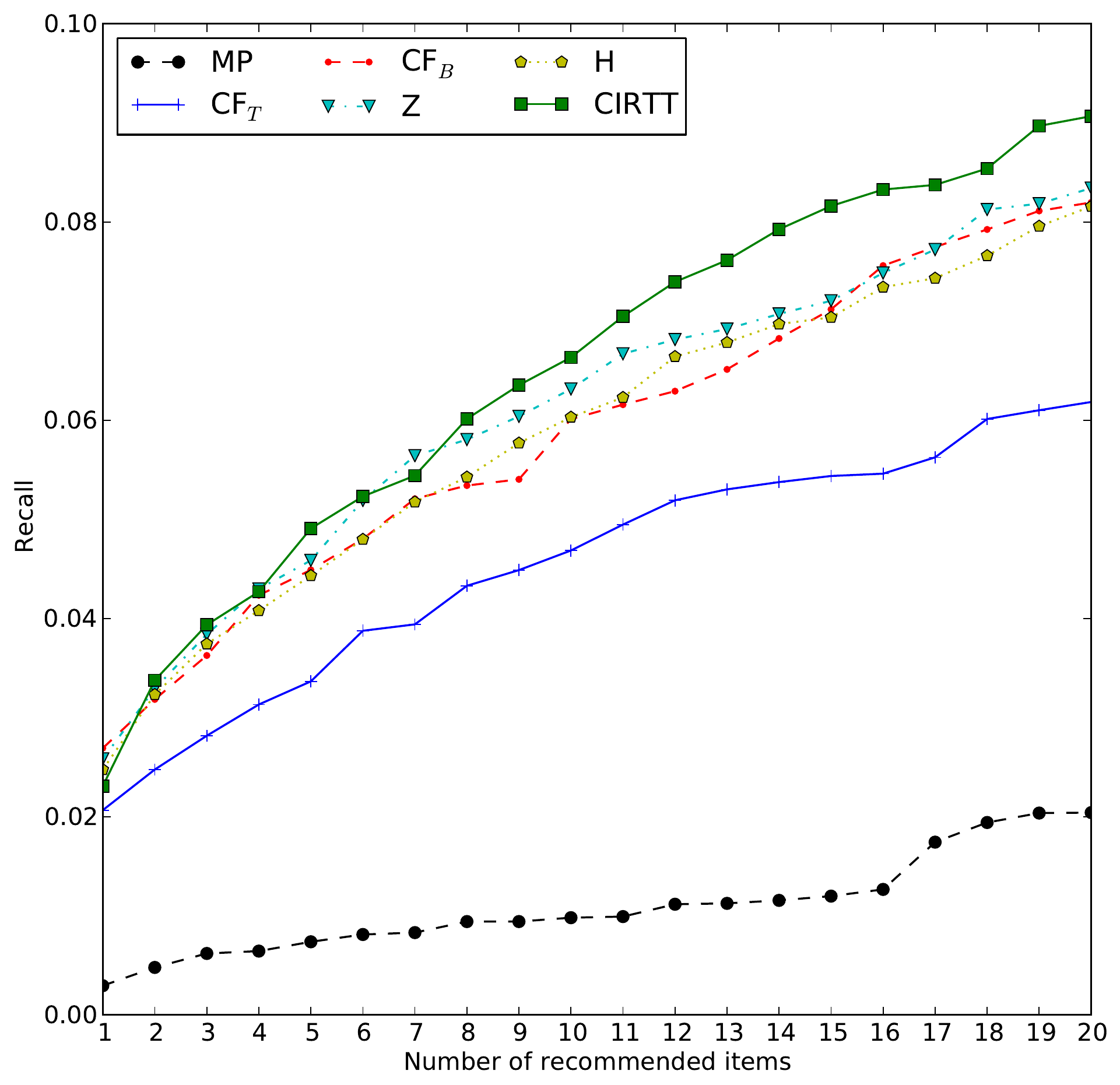}%
		 } 
		 \subfloat[\hspace*{1.8em} Recall \newline CiteULike]{ 
				\includegraphics[width=0.33\textwidth]{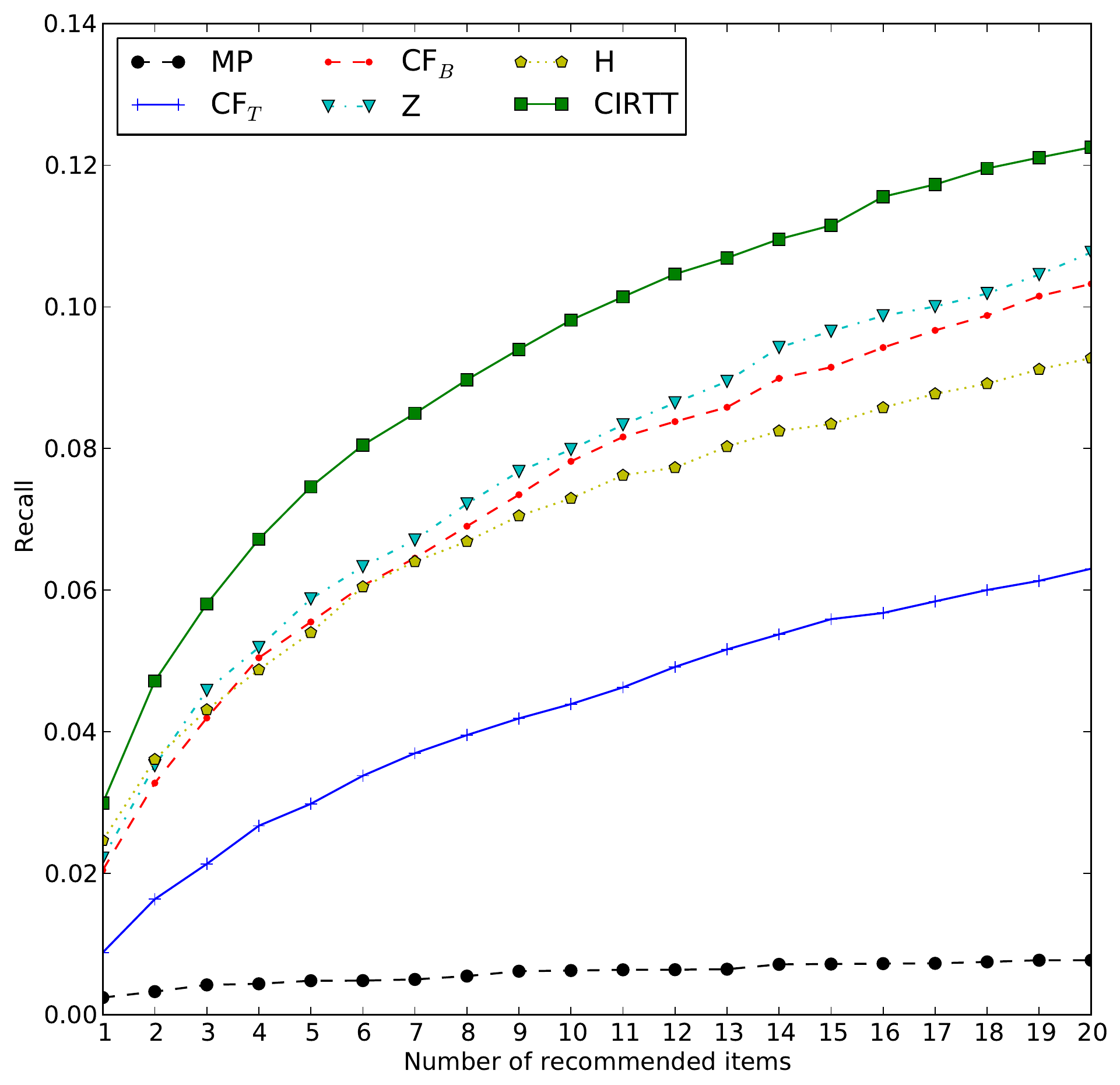}%
		 }
		 \subfloat[\hspace*{1.8em} Recall \newline MovieLens]{ 
				\includegraphics[width=0.33\textwidth]{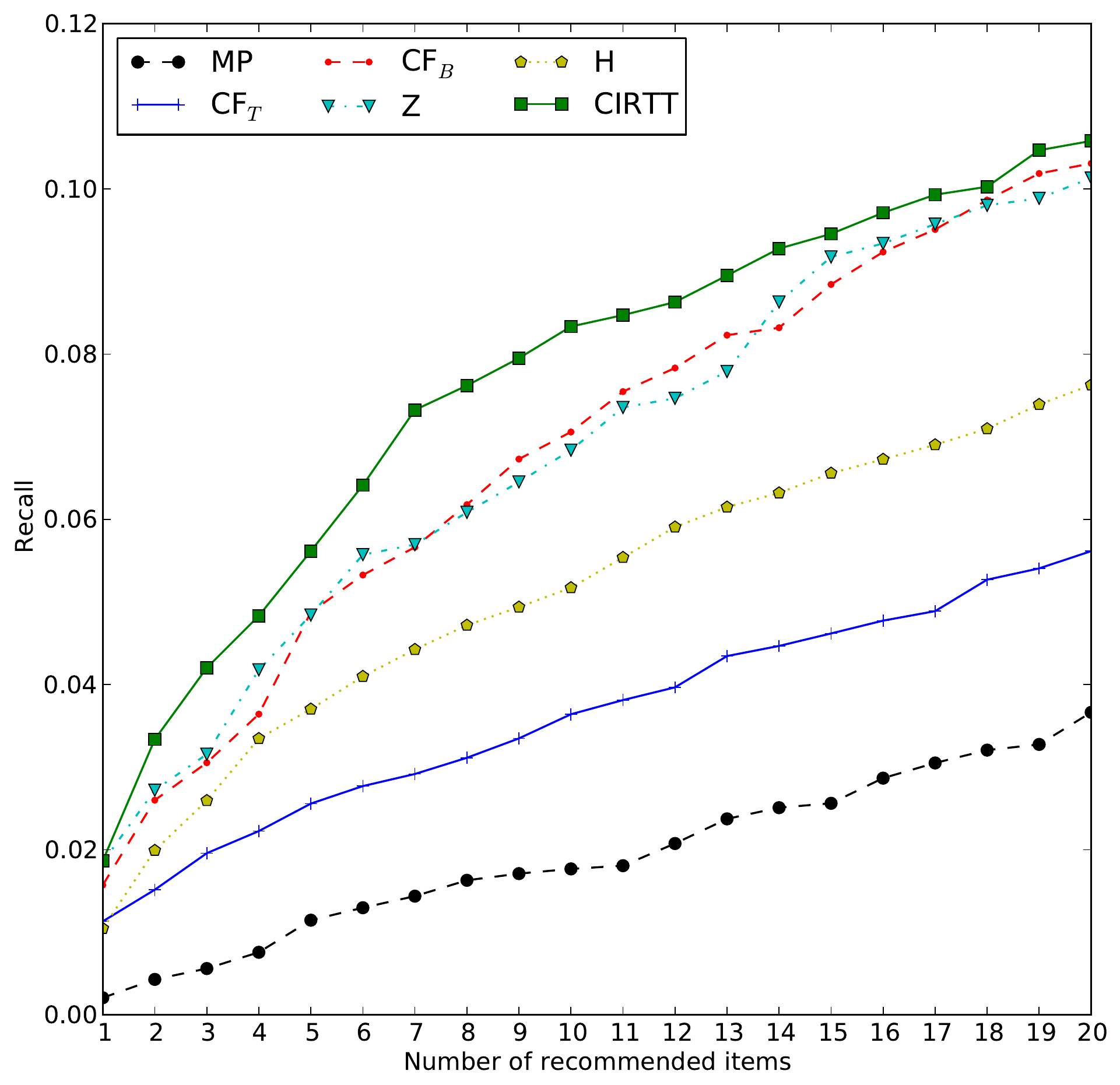}%
		 } 
     \caption{nDCG, MAP and Recall plots for BibSonomy, CiteULike and MovieLens showing the recommendation accuracy of our tag and time based CIRTT approach along with state-of-the-art baseline algorithms for 1 - 20 recommended items (\textit{k}). It can be seen that CIRTT reaches the highest levels of recommender accuracy over all three metrics and on all datasets.}
    \label{fig:ratk}
\end{figure*}

\section{Results}
\label{sec:results}

In this section, we present the results of the evaluation comparing our CIRTT approach to the baseline algorithms described in Section \ref{sec:baselines} with respect to recommender accuracy on three different folksonomy datasets (BibSonomy, CiteULike and MovieLens). 

In an extensive empirical study, Cremonesi et al. \cite{Cremonesi2010} have shown that standard Information Retrieval accuracy metrics (e.g., Recall or nDCG) are well suited to evaluate recommender systems, at least in case of top-$N$ recommendation tasks. Therefore, Table \ref{tab:full_norm} provides measures of accuracy (nDCG@20, MAP@20, R@20) and - additionally - measures of Diversity (D) and User Coverage (UC) for each approach and for each of the three datasets. 

As expected, the MP baseline approach, which is not personalized at all, resulted in the lowest accuracy estimates. Regarding the two traditional CF approaches, the $CF_{B}$ approach, which constructs a binary user-item matrix based on bookmarks, performs better than $CF_{T}$, which is based solely on the user tag-profiles. Regarding the two alternative tag- and time-based approaches, a same phenomenon can be observed as the algorithm of Zheng et al. (Z) \cite{Zheng2011}, that is also based on the binary user-item matrix, performs better than the approach of Huang et al. (H) \cite{Huang2014}, that is based on the user tag-profiles.

With respect to all accuracy metrics (nDCG@20, MAP@20, R@20), our CIRTT approach, that integrates tag and time information using the BLL-equation, performs best in all three datasets (BibSonomy, CiteULike and MovieLens). This may suggest that applying a power-law function as it is done via the BLL-equation is more appropriate to account for effects of recency than an exponential function (Zheng et al. \cite{Zheng2011}) or a linear function (Huang et al. \cite{Huang2014}).  A same pattern of results can be observed when looking at Figure \ref{fig:ratk} that reveals estimates of the nDCG, MAP and Recall measures for different sizes of the recommended item set. We have also tried to integrate the exponential recency function of Zheng et al. in our approach which resulted in lower accuracy estimates than the BLL power law forgetting function.

When looking at the other two not accuracy-based metrics, interestingly, the approach of Huang et al. (H) always results in the lowest Diversity (D) of recommended items. This result might appear because this approach is based on the user tag-profiles and the Diversity metric is calculated based on tags. Finally, as all personalized approaches utilize a user-based CF approach for finding similar users, the measure of User Coverage (UC) does not appear to deviate between the different algorithms. We observed the maximum deviation of 2.53\% within the MovieLens dataset.

\section{Conclusions \& Future Work} \label{sec:con}
In this work we have presented preliminary results of a novel recommendation approach called \textit{Collaborative Item Ranking Using Tag and Time Information (CIRTT)} that aims at improving Collaborative Filtering in social tagging systems. Our algorithm follows a two-step approach as also done in \cite{Huang2014}, where in the first step a potentially interesting candidate item set is found performing user-based CF and in the second step this candidate item set is ranked performing item-based CF. Within this ranking step we integrate the information of frequency and recency of tag use applying the Base-Level Learning (BLL) equation \cite{Anderson2004}. Thus, in contrast to existing approaches that also consider information about tags and time (e.g., \cite{Zheng2011,Huang2014}), CIRTT draws on an empirically well established formalism modeling the reuse probability of memory items (tags) in form of a power-law forgetting function. In recent work, the same formalism has turned out to substantially improve the ranking and recommendation of tags (\cite{domi2014}).

The current evaluation conducted on datasets gathered from three social tagging systems (BibSonomy, CiteULike and MovieLens) reveals that applying the BLL equation also helps to improve the ranking and recommendation process of items. Most important, the results speak in favor of an integrative research endeavor that places a data-driven approach on a theoretical foundation provided by research on human cognition and semiotics.

Our future work will aim at improving the approach presented in this paper. For example, we will examine as to whether the BLL equation can also help to improve the calculation of user similarities and thus, to find more suitable user neighborhoods and candidate items. Additionally, we will put more emphasis on semiotic dynamics that have been found to play out in tagging systems (e.g., \cite{steels2006semiotic}) and how individual learning and forgetting processes are influenced by other individuals' behavior in the system. Moreover, we also plan to further improve the item ranking process using insights of relevant research dealing with recommender novelty and diversity (e.g., \cite{vargas2011rank} in order to increase the user acceptance.

\textbf{Acknowledgments:}
This work is supported by the Know-Center, the EU funded project Learning Layers (Grant Nr. 318209) and the Austrian Science Fund (FWF): P 25593-G22.
The Know-Center is funded within the Austrian COMET Program - Competence Centers for Excellent Technologies - under the auspices of the Austrian Ministry of Transport, Innovation and Technology, the Austrian Ministry of Economics and Labor and by the State of Styria. COMET is managed by the Austrian Research Promotion Agency (FFG).

\balance
\bibliographystyle{abbrv}

\bibliography{rsweb2014}

\end{document}